\documentstyle[manuscript,aps,prbbib]{revtex}

\tighten
\newcommand{\be}{\begin{equation}}
\newcommand{\ee}{\end{equation}}
\newcommand{\bea}{\begin{eqnarray}}
\newcommand{\eea}{\end{eqnarray}}
\newcommand{\bd}{\begin{displaymath}}
\newcommand{\ed}{\end{displaymath}}
\newcommand{\bi}{\begin{itemize}}
\newcommand{\ei}{\end{itemize}}
\newcommand{\bc}{\begin{center}}
\newcommand{\ec}{\end{center}}
\newcommand{\bfl}{\begin{flushleft}}
\newcommand{\efl}{\end{flushleft}}
\newcommand{\bfr}{\begin{flushright}}
\newcommand{\efr}{\end{flushright}}

\def\6{\partial} \def\a{\alpha} \def\b{\beta}
\def\g{\gamma} \def\d{\delta}  \def\e{\epsilon}
  \def\th{\theta}
  \def\l{\lambda}
\def\m{\mu} \def\n{\nu} \def\x{\xi} 
\def\r{\rho} \def\ss{\sigma} 

\def\o{\omega} \def\G{\Gamma}

\def\={\!\!\!&=&\!\!\!}
\def\+{\!\!\!&&\!\!\!+~}
\def\-{\!\!\!&&\!\!\! -~}

\newcommand{\CC}{{\cal C}}
\newcommand{\DD}{{\cal D}}

\newcommand{\TT}{{\cal T}}

\begin{document}

\title{On Local Constraints of D=4 Euclidean Supergravity in Terms of
Dirac Eigenvalues}
\author{
N. Pauna \\ Department of Theoretical Physics\\ Babes-Bolyai
University of Cluj\\ Str. M. Kogalniceanu Nr.1, RO-3400 Cluj,
Romania\\ and\\ Ion V. Vancea \footnote{On leave from Dept. of
Theor. Physics, Babes-Bolyai University of Cluj, Romania}\\
Departamento de F\'{i}sica Te\'{o}rica, Instituto de F\'{i}sica\\
Universidade do Estado do Rio de Janeiro\\ R. S\~{a}o Francisco
Xavier 524, 20550-013 Maracan\~{a}\\ Rio de Janeiro, Brazil}
\address{}
\date{\today}
\maketitle

\begin{abstract}
It has been recently shown that in order to have Dirac eigenvalues
as observables of Euclidean supergravity, certain constraints
should be imposed on the covariant phase space as well as on Dirac
eigenspinors. We investigate the relationships among the
constraints in the first set and argue that these relationships
are not linear. We also derive a set of equations that should be
satisfied by some  arbitrary  functions that enter as coefficients
in the equation expressing the linear dependency of the
constraints in order that the second set of constraints be
linearly independent.
\end{abstract}

It has been recently shown that Dirac eigenvalues can be used as
observables of D=4 Euclidean gravity on compact spacetime without
boundary \cite{lr,lpr1}. This result is based on previous works done in
the framework  of the noncommutative geometry as an underlying
structure of gravity beyond the Planck scale \cite{ncg}. To extend
this idea to supergravity, it is necessary to impose some
constraints, called primaries, on the set of Dirac eigenspinors
\cite{viv2}. These constraints allow us to interpret Dirac
eigenvalues as local observables of N=1 D=4 Euclidean
supergravity. To promote them to global observables, one has to
impose the compatibility of the geometrical structure of the
spacetime with the two sets of constraints. That results in severe
restrictions on spacetime manifolds that admit this type of global
observables \cite{viv2}. However, we should note that this discussion
is related to the problem of realizing the Dirac operator in curved
space. At present, no satisfactory answer to this problem is known,
even though there are many studies on the dirac operator that rely on
different coordinates in spacetime. It is not our purpose to solve this
problem here which is clearly a nontrivial one.

The aim of this letter is to discuss the relationships among the
primaries as well as the secondaries. The basic motivation for
this relies on the fact that all of the previous analysis of this
system is done at the classical level, while the main reason for
introducing this kind of description of Euclidean supergravity
aims at giving insights in the quantum theory.This amounts to
applying one of the quantization methods based on the BRST
symmetry, like BV-BRST or BFV-BRST, which are the most powerful
methods for quantizing the constrained systems \cite{teithen}.
However, as was noticed previously and as will result from the
present discussion, this is a difficult task. Therefore, in this
paper we limit our analysis to the first essential step of the
quantization of the system, namely to the discussion of the
nontrivial relationships among the constraints.

Let us consider N=1 D=4 Euclidean supergravity on a compact (spin)
manifold without boundary. The on-shell supergraviton is given by
the vierbein fields $e^{a}_{\m}(x)$, where $ \m = 1, 2, 3, 4$ and
the gravitino fields $\psi^{\a}_{\m}$ where $\a$ is an index for
an $SO(4)$ spinor that satisfies $\bar{\psi} = \psi^T C$
\cite{spin}. This is our 'Majorana' spinor in the case of
Euclidean supergravity used because $SO(4)$ does not admit an
usual Majorana representation. We can also work with symplectic
spinors \cite{avp}. The gauge group is given by four dimensional
diffeomorphisms, local $SO(4)$ rotations and N=1 local
supersymmetry and its action on the supergraviton is given by \bea
\d e^{a}_{\m} &=& \xi^{\n} \6_{\n} e^{a}_{\m} + \th^{ab} e_{b\m} +
\frac{1}{2}\bar{\e}\g^{a}\psi_{\m}\\ \d \psi^{\a}_{\m} &=&
\xi^{\n} \6_{\n} \psi^{\a}_{\m} + \th^{ab} (\ss_{ab})^{\a}_{\b}
\psi_{\m}^{\b} + \DD_{\m}\e^{\a} \label{gaugetr} \eea where $\xi =
\xi^{\nu}\6_{\n}$ is an infinitesimal vector field on $M$,
$\th_{ab}=-\th_{ba}$ parametrize an infinitesimal rotation and
$\epsilon$ is an infinitesimal Majorana spinor field. Here
$\DD_{\m}$ is the nonminimal covariant derivative acting on
spinors, associated to the usual minimal one. The covariant phase
space is defined to be the space of the solutions of the equations
of motion modulo the gauge transformations. Then the observables
of the theory are functions on the phase space.

In the presence of local supersymmetry, the Dirac operator is
given by
\be
D = i\g^a e_{a}^{\m} (\6_{\m} +
\frac{1}{2} \o_{\m bc}(e,\psi ) \ss^{bc} ), \label{dirop}
\ee
where $\ss^{bc} = \frac{1}{4}[\g^a , \g^b ]$,
$\g^a $'s form a representation of Clifford algebra
$\{ \g^a, \g^b \} = 2\delta^{ab}$ and ${\o}_{\m bc}(e,\psi )$
are the components of spin connection in the presence of local supersymmetry.
The Dirac operator is a first order elliptic operator on $M$. Thus, since
$M$ is compact, $D$ has a discrete spectrum
\be
D \chi^n = \l^n \chi^n ,\label{diraceq} \ee where $ n=0,1,2,
\ldots$. The eigenvalues $\l^n$'s are functions on the space of
all supermultiplets $(e, \psi )$. In order to be interpreted as
observables, they must also be gauge invariant. The invariance of
$\l^n $'s under diffeomorphisms, local $SO(4)$ rotations and local
supersymmetry, is expressed by the following set of equations
\cite{viv1}
\be
{\TT}^{n \m}_{a}{\6}_{\n}e^{a}_{\m}
 -{\Gamma}^{n \m}_{\alpha}{\6}_{\n}{\psi}^{\alpha}_{\m}=0, \label{firstc}
\ee
\be
  {\TT}^{n \m}_{a}e_{b \m}+{\Gamma}^{n \m}{\sigma}_{ab}{\psi}_{\m}=0, \label
 {secondc}
\ee
\be
  {\TT}^{n \m}_{a}\bar{\epsilon}{\gamma}^{a}{\psi}_{\m}
  +{\Gamma}^{n \m}{\DD}_{\m}{\epsilon}=0, \label{thirdc}
\ee
where
\be
{\TT}^{n}_{\m}= \langle {\chi}^{n}|(\frac{\delta}{{\delta}e^{a}_{\m} }D)|
{\chi}^{n} \rangle ~~,~~
{\Gamma}^{n \m}_{\alpha}= \langle {\chi}^{n} | (\frac{\delta}{{\delta}
{\psi}^{\alpha}_{\m}}D) | {\chi}^{n} \rangle ,
\label{functder}
\ee
are the functional derivatives of the Dirac operator with respect to the
graviton and gravitino, respectively. The scalar product is naturally
defined in the spinor bundle on $M$
\be
<{\psi},{\chi}>= \int \sqrt{g}{\psi}^{\ast}{\chi}.
\label{scalp}
\ee
Consistency requires that the following constraints be imposed on the set
of Dirac eigenspinors \cite{viv1}
\be
\{ [b^{\m}(\x ) - c(\l \x )^{\m} ]\6_{\m} + f(\x ) \} \chi^n = 0,
\label{firstcc}
\ee
\be
[\th_{a}^{a}D -g(\th ) +h(\th ) ]\chi^n = 0.
\label{secondcc}
\ee
\be
[j^{\m}_{a} (\e ) \6_{\m} + k_{a} (\e ) +l_{a} ]\chi^n =0
\label{thirdcc} \ee where the following shorthand notations have
been employed \bea b^{\m}(\x )  &=& i \g^{a} b_{a}^{\m}(\x )~~,~~
b_{a}^{\m}(\x ) = \x^{\n}\6_{\n}e_{a}^{\m}
-e_{a}^{\n}\6_{\n}\x^{\m} -2e_{a}^{\n}\x^{\m}\o_{\n
bc}\ss^{bc}\nonumber\\ c(\l ,\x )^{\m} &=& (\l^n - D)\x^{\m}~~,~~
f(\x ) = i\g^{a}\x^{\n}\6_{\n}(e_{a}^{\m}\o_{\m
bc})\ss^{bc}\nonumber \\ c(\l ,\x )^{\m} &=& (\l^n - D)\x^{\m}
~~,~~ g(\th ) = [\g^c e_{c}^{\m}([{\bf \th \ss},\o_{\m ab}] -
\6_{\m}{\bf \th \ss }M_{ab})]\ss^{ab} \nonumber\\ h(\th ) &=&  i
(\l^n - D) {\bf \th \ss}~~,~~ j^{\m }_{a} (\e )  =
\frac{1}{2}\g_{a}\bar{\e}\psi^{\m} ~, ~~~k_{a}(\e ) =
\frac{1}{2}\g_{a}\bar{\e}\psi^{\m}\o_{\m cd}\ss^{cd} \nonumber\\
l_a &=&e_{a}^{\m}[A_{\m cd} - \frac{1}{2}e_{\m d}A_{ec}^{e} +
\frac{1}{2}e_{\m c}A_{ed}^{e}]\ss^{cd} ~~,~~ A_{a}^{\m \n} =
\bar{\e}\g_{5}\g_{a}\DD_{\l}\psi_{\r}\e^{\n \m \l \r}. \label{not}
\eea

The first set of constraints, (\ref{firstc}), (\ref{secondc}) and
(\ref{thirdc}), also called primaries, should be imposed on the
supergravitons. Equations (\ref{firstcc}), (\ref{secondcc}) and
(\ref{thirdcc}) follow as a consequence of the primaries, and
therefore are called secondaries. As was note in \cite{viv2},
primaries as well as secondaries should be taken into account when
the BRST quantization of the system is performed. For example,
both of the sets of constraints determine the partition function
in the path integral approach. Therefore, it is crucial to
elucidate the reducibility of this system. To this end we have to
address the question of linear dependency of the two sets of
constraints.

Let us begin by discussing the relationships among the primaries.
To simplify the expressions in what follows, let us denote the
constraints (\ref{firstc}), (\ref{secondc}) and (\ref{thirdc}) by
${\Sigma}^{n}_{\n}(\TT,\Gamma)=0$,
${\Theta}^{n}_{ab}(\TT,\Gamma)=0$ and ${\Phi}^{n}(\TT,\Gamma)=0$,
respectively. Now, if we multiply ${\Theta}^{n}_{ab}$ at left by
$\bar{\epsilon}{\gamma}^{a}$ and sum over $a$, and then multiply
the result at right by $e^{b}_{\rho}{\psi}^{\rho}$, we come across
the first term in ${\Phi}^{n}$. Thus we can immediately write down
a relationship between the second and the third primary
\be
{\Phi}^{n}-\bar{\epsilon}{\gamma}^{a}{\Theta}^{n}_{ab}
{\psi}^{b}={\Gamma}^{n
\m}\DD_{\m}{\epsilon}-\bar{\epsilon}{\gamma}^{a} {\Gamma}^{n
\n}{\sigma}_{ab}{\psi}_{\m}{\psi}^{b} \label{firstr} \ee Eq.
(\ref{firstc}) does not express, in general, the linear
independency of ${\Theta}^{n}_{ab}$ and ${\Phi}^{n}$. This is true
only if the arbitrary spinor $\e$ obeys the following equation
\be
\DD_{\m}{\epsilon}=\bar{\epsilon}{\gamma}^{a}
{\Gamma}^{n \n}{\sigma}_{ab}{\psi}_{\m}{\psi}^{b},
\label{spineq}
\ee
which selects from general local supersymmetry transformations a certain
class
of transformations of which parameters obey (\ref{spineq}). We can also
see that
there is no other linear relationship among these constraints based on
the second
terms in ${\Theta}^{n}_{ab}$ and ${\Phi}^{n}$ and therefore we conclude that
these two constraints are not linearly independent.

In a similar manner we can approach the relationship between
${\Sigma}^{n}_{\n}$ and ${\Theta}^{n}_{ab}$. In this case, we see
that ${\Sigma}^{n}_{\n}$ involves the derivatives of the
supergraviton, while ${\Theta}^{n}_{ab}$ involves only its
components, besides, of course, the ${\TT}^{n \m}_{a}$ and
${\Gamma}^{n \m}_{\alpha}$ terms which both involve integrals of
supergraviton and its derivatives. Therefore, we cannot obtain a
linear relationship between these two constraints. However, we can
obtain, after some simple algebra, the following equation
\be
{\Sigma}^{n}_{\n} - \6_{\n}{\Theta}^{n}_{ab} = \6_{\n}(\Gamma^{n
\m}_{\a} (\ss^{a}_{a} )^{\a}_{\b} \psi^{\b}_{\m}) - \Gamma^{n
\m}_{\a} \6_{\n} \psi^{\a}_{\m} - \6_{\n} \TT^{n \m}_{a}e^{a}_{\m
}. \label{secondr} \ee In the right-hand side of
Eq.(\ref{secondr}) we have written down explicitely the spinor
index in the first term. This shows us that another relationship
based on the identification of the last two terms, modulo some
functions, is not possible.

Because the relationship between $\Phi^n $ and $\Theta^{n}_{ab}$
implies the derivation of the later and because (\ref{firstr})
expresses the relationship between $\Phi^n $ and
$\Theta^{n}_{ab}$, we see that a relationship between $\Phi^n$ and
$\Sigma^{n}_{\n }$ would imply the derivatives of $\Phi^n$. A
final relationship among all the constraints and thus between
$\Phi^n$ and $\Sigma^{n}_{\n}$ too, is also nonlinear and is given
by \bea \6_{\n}\bar{\e}\g^a \Theta^{n}_{ab}\psi^b + \bar{\e}\g^a
\Theta^{n}_{ab}\6_{\n}\psi^{b} + \sum_{a\neq b}\bar{\e}\g^a
\6_{\n} \Theta^{n}_{ab}\psi^b - \6_{\n}\Phi^n + \sum_a
\bar{\e}\g^a \sigma^{n}_{\n}\psi_a = \nonumber\\ \sum_a
(\bar{\e}\g^a \6_{\n} (\G^{n\n}\ss_{aa}\psi_{\m}) +
\G^{n\m}\6_{\n}\psi_{\m} + \6_{\n}\TT^{n\m}_{a} e_{a\m})\psi^a -
-\6_{\n}(\G^{n\m }\DD_{\m}{\e} - \bar{\e}\g^a\G^{n\n}\ss_{ab}\psi_{\n}
\psi^b ). \label{thirdr}
\eea

Some comments are in order now. As we can see from (\ref{firstr}),
(\ref{secondr}) and (\ref{thirdr}), the set of primaries is not
linearly independent. However, as we have already noticed, for a
particular set of local supersymmetries given by the solutions of
(\ref{spineq}), (\ref{firstr}) turn into a linear equation among
the three primaries, where the third one appears with a null
coefficient. The functions multiplying $\Phi^n$ and
$\Theta^{n}_{ab}$ are the ones in the left-hand side of
(\ref{firstc}). This has some important consequences in the BRST
quantization. Indeed, in the partition function of Fadeev-Popov
quantization method, the constraints
${\Sigma}^{n}_{\n}(\TT,\Gamma)=0$,
${\Theta}^{n}_{ab}(\TT,\Gamma)=0$ and ${\Phi}^{n}(\TT,\Gamma)=0$
enter the exponential of the action through a term of the
following form
\be
S_{c} = {\int}({\sigma}^{\n}_{n}{\Sigma}^{n}_{\n}+
{\tau}^{ab}_{n}{\Theta}^{n}_{ab}+{\phi}_{n}{\Phi}^{n}), \label{sconstr}
\ee
where ${\sigma}^{\n}_{n}$, ${\tau}^{ab}_{n}$ and ${\phi}_{n}$ associated
to the gauge averaging conditions. The corresponding ghosts, denoted by
$s^{\n}_{v}$, $t^{ab}_{n}$ and $f_{n}$ enter the exponential through the
following action
\be
 S_{gh}={\int}(s^{\n}_{v}{\delta}_{\alpha}{\Sigma}^{n}_{\n}
 +t^{ab}_{n}{\delta}_{\alpha}{\theta}^{n}_{ab}
 +f_{n}{\delta}_{\alpha}{\phi}^{n})c^{\alpha}, \label{sghosts}
\ee where $c^{\a}$ are the ghosts associated to the gauge
transformations, generally denoted by $\d_{\a}$. Now, if for
certain supersymmetry transformations, i. e. those for which the
parameters satisfy (\ref{spineq}), the constraints become linearly
independent, the theory, irreducible up to now, becomes a first
reducible theory. Thus we obtain an enhancement of the content of
ghost-antighost fields. If we denote the linear constraint by $\CC
(\Sigma^{n}_{\n}, \Theta^{n}_{ab}, \Phi^n ) =0$ we have the
corresponding ghost-antighost structure associated to this
equation \cite{teithen}.

Another issue is what the relationships among the secondaries are.
The secondaries restrict the set of eigenspinors and implicitely
the possible $\TT^{n\m}_{a}$'s and $\G^{n\m}_{\a}$'s,
respectively, which are diagonal matrix elements on eigenspinors.
Therefore, taking into account the secondaries, the number of
${\Sigma}^{n}_{\n}$'s, ${\Theta}^{n}_{ab}$'s and ${\Phi}^{n}$'s in
(\ref{sconstr}) and (\ref{sghosts}) should reduce. Thus we can see
that investigating the dependence of secondaries is important even
if these constraints are directly imposed on the covariant phase
space.

Let us denote (\ref{firstcc}), (\ref{secondcc}) and
(\ref{thirdcc}) by $\CC_1\chi^n =0$, $\CC_2 \chi^n =0$ and $\CC_3
\chi^n =0$, respectively. We are looking for a linear relationship
among all of the secondaries. Now if we consider $\CC_i$'s as some
algebraic function of even Grassman parity, we obtain after some
simple algebra the following equations
\be
e^{\r}_c j^{\n}_d (b^{\m} +c^{\m})f_{1\r}f^{d}_{1\n} + (b^{\n}
+c^{\n})j^{\r}_d e^{\m}_c f_{2\n}f^{d}_{2\r} + (b^{\n}
+c^{\n})e^{\r}_{c}j^{\m}_d f_{3\n}f_{3\r}f^{d}_{3} =0
\label{firstrrr} \ee and \bea i\g^c \theta^{a}_{a}(e^{\r}_c
j^{\n}_d f_{1\r}f_{1\n}f +(b^{\n} +c^{\n}) j^{\r}_d
\theta^{a}_{a}\g^c e^{\m}_c \o_{\m ef}\ss^{ef}f_{2\n}f^{d}_{2\r} +
\nonumber\\ (b^{\n} +c^{\n})e^{\r}_c (k_d + l_d
)f_{3\n}f_{3\r}f^{d}_{3}) +(h-g)f_{2\n}f^{d}_{2\r} = 0
\label{secondrrr} \eea where
$f_{1\r}$,$f^{d}_{1\n}$,$f_{2\n}$,$f^{d}_{2\r}$,$f_{3\n}$,$f_{3\r}$
and $f^{d}_{3}$ are the coefficients of the constraints in the
equation that describes their linear dependency. These
coefficients are in number of 32 of them and should obey the above
equations. Because the number of functions exceeds that of
equations, the system is not well determined. However, some
particular solutions could be find in principle by fixing 30 of
the functions, let say to some constants. Eqs. (\ref{firstrrr})
and (\ref{secondrrr}) depend on the gauge transformations. They
also include some spinorial objects and thus are not trivial. If
they admit nontrivial solutions then there is a linear
relationship among all of the secondaries.

Now let us take two of the secondaries. A linear combination of $\CC_1$
and $\CC_2$ implies that the following equations hold
\be
e^{\n}_c g_{1\n} (b^{\m} + c^{\m}) +(b^{\n} + c^{\n})g_{2\n}e^{\m}_c =0
\label{thirdrrr}
\ee
and
\be
i\g^c \theta^{a}_{a}(e^{\n}_{c}g_{1\n}f + (b^{\n} +c^{\n})g_{2\n}
e^{\m}_c \o_{\m de} \ss^{de}) - (b^{\n} + c^{\n})g_{2\n}(g-h) =0 .
\label{fourthrrr} \ee Eqs. (\ref{thirdrrr}) and (\ref{fourthrrr})
must be satisfied by the two unknown spacetime vectorfields
$g_{1\n}$ and $\g_{2\n}$,respectively, which are now the
coefficients in the linear equation among constraints.  Similarly,
if we take $\CC_2$ and $\CC_3$, another set of two equations must
hold, namely
\be
j^{\n}_{a} g_{3\n} e^{\m}_c + e^{\n}_{c}g^{a}_{4\n}j^{\m}_{a} = 0
\label{fifthrrr}
\ee
and
\be
j^{\n}_a g_{3\n} (i\theta ^{a}_{a}\g^c e^{\m}_c\o_{\m de}\ss^{de} -g + h )
+ i\theta^{b}_{b} \g^c e^{\n}_{c} g^{a}_{4\n}(k_a + l_a ) = 0,
\label{sixthrrr}
\ee
where the unknown spacetime vectorfield is $g_{3\n}$, while $g^{a}_{4\n}$
is a double index unknown quantity. The final possible linear relationship
holds if the following equations hold
\be
j^{\n}_a g^{a}_{5\n} (b^{\m} + c^{\m}) + (b^{\n} +
c^{\n})g^{a}_{6\n}j^{\m}_{a} = 0 \label{seventhrrr}
\ee
and
\be
j^{\n}_{a}g^{a}_{5\n}f + (b^{\n} + c^{\n})g^{a}_{6\n}(k_a + l_a ) =0,
\label{eightrrr}
\ee
where $g^{a}_{5\n}$ and $g^{a}_{6\n}$ are two indices unknown objects.

Let us make some brief comments on the equations (\ref{thirdrrr})
- (\ref{eightrrr}). The unknown objects that must satisfy these
equations represent a set of 60 unknown coefficients. In the most
general case, the equations group as shown above. Then for each
set of equations we have a redundant number of variables. To find
some particular solutions we can fix some of them in an arbitrary
way. Another fixing procedure is to consider the whole set
equations as a system and then to require that some of the objects
entering one of the subsets of equations be identically with some
of the objects entering another subset, e. g. $g^{a}_{4\n} =
g^{a}_{5\n}$. This amounts to impose the existence of linear
relationships among {\em all} of the secondaries simultaneously.
If the equations (\ref{firstrrr})-(\ref{eightrrr}) have nontrivial
solutions, the secondaries are not linear independent. As a
consequence, follows the enlarging of the set of the allowed
eigenspinors and implicitely the enlarging of the set of diagonal
matrices in (\ref{sconstr}) and (\ref{sghosts}), respectively. We
should remark that the equations derived above are not trivial and
not easy to be solved in the general case. This postpone the BRST
analysis of the system to some future works. We note that since
local supersymmetry and $SO(4)$ invariance in supergravity
coupled to matter normally require field-depended transformations
for consistency, the gauge-fixing must be carried out in accordance
with variations of gravitino and graviton. Their consistency with the
global symmetries of the theory must be explicitely checked.

To conclude, we have analyzed in this letter the dependency of the
primaries and secondaries of the Euclidean supergravity in terms
of Dirac eigenvalues. We have shown that in general the primaries
are not linearly independent and thus the theory is an irreducible
one in the BRST language. However, for some local supersymmetry
transformations, it is possible that the theory become first stage
reducible and that the ghost-antighost structure be enlarged. We
have also derived the set of equations
(\ref{firstrrr})-(\ref{eightrrr}) which impose the linear
independency of secondaries. As a consequence of these equations,
the set of admissible eigenspinors is enlarged, too, and that, at
its turn, modifies the extended action that enters the partition
function of the system.

\acknowledgements{ The authors would like to thank to W. Troost,
A. Van Proeyen and J. Gomis for discussions. I. V. Vancea would also like to
thank to J. Helayel-Neto for hospitality at Centro Brasiliero de Pesquisas
Fisicas where part of this work was acomplished and to acknowledge
FAPERJ for a postdoc fellowship.}


\begin{references}
\bibitem{lr}G. Landi and C. Rovelli, Phys. Rev. Lett.78, 3051-3054(1997)
\bibitem{lpr1}G. Landi and C. Rovelli, Mod. Phys. Lett. A13(1998)479-494
\bibitem{ncg}For reviews and references, see:
A. Connes, {\it Noncommutative Geometry }, (Academic Press,
New York, 1994); J. Madore, {\it An Introduction to Noncommutative
Differential Geometry and Its physical Applications}, LMS Lecture Notes 206,
(1995); J. Fr\"{o}hlich, O. Grandjean and A. Recknagel, hep-th/9706132;
G. Landi, {\it An Introduction to Noncommutative Spaces and Their Geometries},
(Springer-Verlag, 1998)
\bibitem{viv1}I. V. Vancea, Phys. Rev. Lett. 79, 3121-3124(1997);
Erratum-ibid.80, 1355(1998)
\bibitem{viv2}I. V. Vancea, {\it Euclidean Supergravity in Terms of Dirac
Eigenvalues}, e-print gr-qc/9710132, Phys. Rev. D58(1998)045005
\bibitem{teithen}C. Teitelboim and M. Henneaux, {\it Quantization of Gauge
System}, (Princeton University Press, 1992)
\bibitem{spin}P. Van Niewvenhuizen,in{\it Relativity, Groups and Topology II},
Proceedings of the Les Houches Summer School, Session XL, edited
by R. Stora and B. S. DeWITT(North-Holland, Amsterdam, 1984)
\bibitem{avp}A. Van Proeyen, {\it Superconformal Algebras} in {\it
Super Field Theories}, eds. H. C. Lee and al., (Plenum Press), 547
(1987); A. Van Proeyen, in {\it 1995 Summer School in High Energy
Physics and Cosmology}, The ICTP series in theoretical physics -
vol.12(World Scientific, 1997), eds. E. Gava et al.,
hep-th/9512139


\end{references}
\end{document}